\documentclass[12pt]{iopart}

\usepackage{epsfig}

\begin{document}
%\title[Bicritical point in an out-of-equilibrium Ising model]
%{Bicritical point in an out-of-equilibrium Ising model}
\title[Nonequilibrium lattice models with two symmetric 
absorbing states] {Critical behavior in lattice models with
two symmetric absorbing states}

\author{\'Attila L. Rodrigues$^1$, Christophe Chatelain$^2$,
T\^ania Tom\'e$^1$ and M\'{a}rio J. de Oliveira$^1$}

\address{$^1$ Instituto de F\'{\i}sica,
Universidade de S\~{a}o Paulo,
Caixa Postal 66318, 
05314-970 S\~{a}o Paulo, S\~{a}o Paulo, Brazil}

\address{$^2$ Groupe de Physique Statistique, 
D\'epartment de Physique de la Mati\`ere et des
Mat\'eriaux, Institut Jean Lamour (CNRS UMR 7198),
Universit\'e de Lorraine Nancy, BP 70239, F-54506 
Vandoeuvre les Nancy Cedex, France}

\begin{abstract}

We analyze nonequilibrium lattice models with up-down symmetry
and two absorbing states by mean-field approximations
and numerical simulations in two and three dimensions.
The phase diagram displays
three phases: paramagnetic, ferromagnetic and absorbing.
The transition line between the first two phases belongs to
the Ising universality class and between the last two, to
the direct percolation universality class. The two lines meet
at the point describing the voter model and
the size $\ell$ of the ferromagnetic phase vanishes with
the distance $\varepsilon$ to the voter point as
$\ell\sim\varepsilon$,
%$\ell\sim\varepsilon^\phi$ where the crossover exponent
%$\phi$ is found to be equal do 1,
with possible logarithm corrections in two dimensions.
%We show that
%the size $\ell$ of the ferromagnetic phase vanishes with
%the distance $\varepsilon$ to the voter point as
%$\ell\sim\varepsilon^\phi$ where $\phi=$ in two dimensions,
%$\phi=$ in three dimensions and $\phi=1$ according to mean-field.

\end{abstract}

\pacs{05.70.Ln, 05.50.+q, 64.60.Ht}

\maketitle

%----------------------------------------------------------
\section{Introduction}

Studies of phase transitions and critical phenomena
in equilibrium and nonequilibrium systems have shown that the
critical behavior can be organized into universality classes,
which are identified by a small number of features among them
the symmetry.
The most widely studied universality
class, due to its experimental implications, is the
Ising universality class which includes equilibrium
as well as nonequilibrium systems \cite{grinstein85,oliveira93}. 
The main feature of the Ising universality
class is the up-down symmetry. Another class,
that includes only nonequilibrium systems, is the
direct percolation (DP) universality class \cite{marro99,henkel08,hinrichsen00}.
The main feature that distinguishes this class from the others is the 
presence of a single absorbing state. 
That is, if a system displays a continuous phase transition to a
single absorbing state with no extra symmetry or conserving laws
then its critical behavior belongs to the DP universality class
\cite{janssen1981,grassberger1982}.

Our purpose here is to analyze models that
possess features pertaining to the two universality classes.
%
%Usually, a system having the features associated to a certain
%universality class will manifest the critical behavior if
%there is at least one parameter whose variation
%will drive the system to criticality.
%If we consider a system having the features of two
%universality classes A and B (which we denote by AB),
%then it should be necessary two parameters to drive the system to
%the critical point belonging to AB.
%If a system with the features AB 
%has just one parameter we expect two separate transition points:
%one belonging to A and another belonging to B.
%If there is a second parameter, then its variation 
%may thus force the two critical points
%to merge into the transition point AB.
%
Here, we focus on models belonging to the Ising
and DP universality classes, that is, systems with
up-down symmetry and an absorbing state
\cite{droz03,hammal05,vasquez08,nam11,park12}.
Due to the up-down symmetry, these systems, in fact, have two
symmetric single absorbing states.
%**************
Lattice models of this type have been studied numerically
\cite{droz03} revealing the existence of two transition points
when a parameter is varied: a symmetry breaking point belonging to the
Ising universality class and another belonging to the DP universality class.
Another numerical study in two
dimensions by the method of Langevin equation \cite{hammal05},
with two parameters, has shown that the two critical points evolve
into two transition lines that meet at a certain point,
given rise to a third transition line. This study confirm the
Ising and DP character of the two lines and the existence
of a point on the third line corresponding to the voter model
\cite{oliveira93}, which is not the meeting point. 
It has been claimed that the
symmetry breaking transition is not of the Ising type
\cite{nam11}, but Monte Carlo simulations
on a two-dimensional interacting monomers model \cite{park12}
confirmed the Ising transition. In the present study, we wish 
%Motivated by this controversial point, the present study wishes
%One of the motivations of the present work is
%clarify, esclarecer, elucidar
to clear up this point and analyze the role played by
the voter model. This is achieved by studying lattice models
with up-down symmetry and two absorbing states in two
and three dimensions by numerical simulations
and mean field approach. Our results confirmed the existence
of the Ising and DP lines
and revealed that two lines meet at the voter point
and that this is the only point at the third line which is critical. 
%***************

\begin{figure}
\centering
\epsfig{file=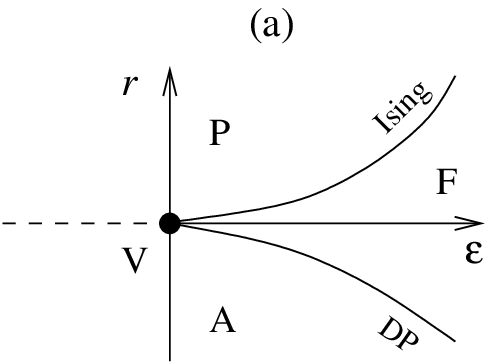,width=6cm}
\hskip1.5cm
\epsfig{file=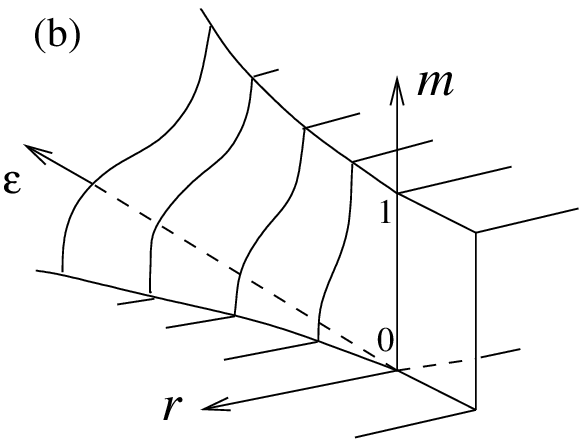,width=6.5cm}
\caption{(a) Phase diagram in the field parameters
$\varepsilon$ and $r$. The full circle, located at the origin,
corresponds to the voter model (V). The phases are:
paramagnetic (P), ferromagnetic (F) and absorbing (A). 
The solid lines are continuous phase transitions belonging
to the Ising and DP universality classes
and the dashed line is a discontinuous phase transition.
(b) Magnetization $m$ as a function of $\varepsilon$ and $r$.
}
\label{cross}
\end{figure}

The phase diagram in the two-dimensional space of parameters
displays three phases: paramagnetic (P), ferromagnetic (F) and
absorbing (A), as shown in figure \ref{cross}a. 
The PF line that separates the P and F phases
is a critical line belonging to the Ising universality class. 
The FA line that separates the A phase and the F phase,
understood as an active phase, is a critical line belonging to
the DP universality class. The two lines, PF and FA, meet
at the point corresponding to the voter model.
As to the PA line, that separates the P and A phases,
it is a discontinuous transition line that ends at the voter point.
Along the PA line the jump in the order parameter is a constant
and equals the jump observed in the voter model.
For an appropriate description of the critical properties 
around the voter point, it is convenient to introduce two
parameters, $\varepsilon$ and $r$, that define a reference frame
centered with origin at the voter point. The $\varepsilon$-axis and $r$-axis
are parallel and perpendicular to the direction defined by
the discontinuous transition line, as shown in figure \ref{cross}a.

%\section{Scaling relations}

The order parameter $m$ of the F phase, shown in figure \ref{cross}b
is assumed to obey the following scaling relation, around the voter point,
\begin{equation}
m = \psi_{\pm}(\varepsilon r^{-1/\phi}),
\end{equation} 
where $\phi$ is the crossover exponent
and $\psi_+$ and $\psi_-$ are valid for $\varepsilon>0$ and
$\varepsilon<0$, respectively. The universal function $\psi_-(x)$
is a step function.
The universal function $\psi_+(x)$ vanishes according to
$\psi_+(x) \sim (a-x)^{\beta_{\rm I}}$, where $\beta_{\rm I}$
is the Ising order parameter exponent and $a$ is a positive constant so that,
near the Ising critical line, the order Ising parameter $m$
behaves as 
\begin{equation}
m = A (\varepsilon_{\rm I} - \varepsilon)^{\beta_{\rm I}},
\end{equation}
where $\varepsilon_{\rm I}=ar^{1/\phi}$ and
the amplitude $A$ diverges as $A \sim r^{-\beta_{\rm I}/\phi}$.
The universal function $\psi_+(x)$ approaches its maximum value 1
according to $1-\psi_+\sim (x+b)^{\beta_{\rm DP}}$
where $\beta_{\rm DP}$
is the DP order parameter exponent and $b$ is a constant so that,
near the DP critical line, the DP order parameter $\rho=1-m$ behaves as 
\begin{equation}
\rho = B (\varepsilon+\varepsilon_{\rm DP})^{\beta_{\rm DP}},
\end{equation}
where $\varepsilon_{\rm DP}=br^{\phi}$ and
the amplitude $B$ diverges as $B \sim r^{-\beta_{\rm DP}/\phi}$.

The Ising and DP critical lines occur when $m\to0$ and $\rho\to0$,
respectively, and are given by $\varepsilon=ar^{1/\phi}$ and
$\varepsilon=-br^{1/\phi}$ so that the size of the F phase
behave near the voter point as
\begin{equation}
\ell \sim \varepsilon^{\,\phi}.
\end{equation}

%-------------
%-----------

%----------------------------------------------------------
\section{Model}

We consider stochastic variables that take two values,
$\sigma_i=\pm 1$, called spin variables.
They are located on the sites of a regular lattice which
is either a two-dimensional triangular lattice or a
three-dimensional cubic lattice. Notice that both lattices
have the same coordination number, which is six.
The system evolves in time according to 
a continuous time Markov process with a single-site change.
The single-site transition rate $w_i(\sigma)$ from
$\sigma=(\sigma_1,\ldots,\sigma_i,\ldots,\sigma_N)$
to $\sigma^i=(\sigma_1,\ldots,-\sigma_i,\ldots,\sigma_N)$
is set up in such way as to hold the properties:
a) it has the up-down symmetry, that is, it is invariant
under the transformation $\sigma\to-\sigma$;
b) it depends on the neighboring spin variables only through the
sum of these spin variables. A transition rate that fulfils these
two properties is given by
\begin{equation}
w_i(\sigma) = \frac12 \{1-\sigma_i f(s_i)\},
\qquad\qquad
s_i=\sum_\delta\sigma_{i+\delta},
\end{equation}
where the sum in $\delta$ extends over the six nearest neighbor sites,
and $f(s)$ is an odd function of $s$.

The most general form of $f(s)$ has three parameters. 
Let us denote by $\sigma_j$, $j=1,2,3,4,5,6$ the spins of the six nearest
neighbor sites of a central site $j=0$, so that
\begin{equation}
s_0 = \sigma_1+\sigma_2+\sigma_3+\sigma_4+\sigma_5+\sigma_6.
\end{equation}
Taking into account that $\sigma_j=\pm1$, then $s_0$ may take 
only the values $0,\pm2,\pm4,\pm6$
%so that $f$ takes the values $0,\pm f_0,\pm f_1,\pm f_2$
from which follows that the most general form of an odd function $f(s)$
can be written as
\begin{equation}
f(s) = \frac{a}{2} s + \frac{b}{8} s^3 + \frac{c}{32} s^5,
\end{equation}
where $a$, $b$ and $c$ are parameters. Here, it is convenient to
define new parameters $p_0=[1+f(6)]/2$, $p_1=[1+f(4)]/2$ and 
$p_2=[1+f(2)]/2$, that is,
\begin{equation}
p_0 = \frac12(1+3a+27b+243c),
\end{equation}
\begin{equation}
p_1 = \frac12(1+2a+8b+32c),
\end{equation}
\begin{equation}
p_2 = \frac12(1+a+b+c).
\end{equation}
In terms of these parameters, the transition probabilities
are as shown in table \ref{trans}.

%------------
\begin{table}
\begin{center}
\caption{\label{trans}
Probability of transition $\sigma_i\to-\sigma_i$. }
\bigskip
\begin{tabular}{|c|c|c|c|c|c|c|c|}
\hline
$\sigma_i$ & -6 & -4 & -2 & 0 & +2 & +4 & +6 \\
\hline
+1 & $p_0$ & $p_1$ & $p_2$ & $1/2$ & $1-p_2$ & $1-p_1$ & $1-p_0$ \\
-1 & $1-p_0$ & $1-p_1$ & $1-p_2$ & $1/2$ & $p_2$ & $p_1$ & $p_0$ \\\hline
\end{tabular}
\end{center}
\end{table}
%----------

Two well known models are special cases. When $p_0=1$, $p_1=5/6$
and $p_2=2/3$, ($a=2/3$, $b=c=0$) we recover the so called voter model
\cite{oliveira93,hammal05}
whose transition rate is given by
\begin{equation}
w_i^{\rm voter}(\sigma) = \frac12 \{1-\frac13\sigma_i s_i\},
\qquad\qquad
s_i=\sum_\delta\sigma_{i+\delta}.
\end{equation}
The Glauber-Ising model, whose transition rate is given by
\begin{equation}
w_i^{\rm GI}(\sigma) = \frac12 \{1-\sigma_i \tanh(Ks_i)\},
\qquad\qquad
s_i=\sum_\delta\sigma_{i+\delta},
\end{equation}
is recovered when the parameter are connected by
\begin{equation}
p_0 = \frac12(1+\tanh 6K),
\end{equation}
\begin{equation}
p_1 = \frac12(1+\tanh 4K),
\end{equation}
\begin{equation}
p_2 = \frac12(1+\tanh 2K).
\end{equation}
In this case, detailed balance is obeyed and
the stationary distribution is the Gibbs distribution
associated to the Ising Hamiltonian.
In other cases there is no detailed balance as can be
seen by the following sequence of transitions. Consider two
nearest neighbor sites initially in states $(++)$.
All the other nearest neighbors of this pair of spins are
positive except one of them which is negative.
Next, they change their states according to the cyclic sequence
$(++)\to(+-)\to(--)\to(-+)\to(++)$.
Using the rules of table \ref{trans}
we find that the ratio between the transition 
probabilities of this path and of its reverse equals
$(1-p_1)^2p_2p_0/(1-p_0)(1-p_2)p_1p_1$, which
in general is distinct from the unit. Thus, there is no
microscopic reversibility and no detailed balance.

To meet the two features of the Ising and DP universality 
class, that is, up-down symmetry and two equivalent single
absorbing states, the transition rate should lead to an absorbing state.
This is accomplished by imposing the
restriction $p_0=1$ because in so doing
the transition rate will vanish whenever all spins are up or
all spins are down, which are the two states
identified as the equivalent absorbing states. 
In this case, the transition rates do not
obey detailed balance, which amounts to say that
the two features mentioned above are only possible in systems
that in the stationary states are out of equilibrium.
When $p_0=1$,
the phase diagram is restricted to the plane $(p_1,p_2)$,
and the voter model corresponds to the point
$(5/6,2/3)$ of this diagram.
As we shall see,
the phase diagram displays a paramagnetic phase,
a ferromagnetic phase and an absorbing state,
as explained before.
%All three phases meet at the voter point, which turn
%out to be a bicritical point.

%----------------------------------------------------------
\section{Mean-field approximation}

A qualitative phase diagram is obtained by the use of
a mean-field approximation. To this end, we start by writing
the function $f(s)$ in the form
\begin{equation}
f(s) = \frac{A}{6} \sum_i\sigma_i
+\frac{B}{20}\sum_{i,j,k}\sigma_i\sigma_j\sigma_k
+\frac{C}{6} \sum_{i,j,k,\ell,n}\sigma_i\sigma_j\sigma_k\sigma_\ell\sigma_n
\end{equation}
where the first summation runs over $i$ from 1 to 6,
and has 6 terms;
the second summation runs over $i,j,k$ with the
restriction $1\leq i<j<k\leq6$ and has 20 terms;
and the third summation runs over $i,j,k,\ell,n$ with the
restriction $1\leq i<j<k<\ell<n\leq6$ and has 6 terms.
The parameters $A$, $B$, and $C$ are related to $p_0$, $p_1$,
and $p_2$ by
\begin{equation}
p_0 = \frac12(1+A+B+C)
\end{equation}
\begin{equation}
p_1 = \frac12(1+\frac23A-\frac23C)
\end{equation}
\begin{equation}
p_2 = \frac12(1+\frac13A-\frac15B+\frac13C)
\end{equation}

The time evolution of the magnetization $\langle \sigma_0\rangle$
is given by
\begin{equation}
\frac{d}{dt}\langle \sigma_0\rangle = -2\langle \sigma_0w_0(\sigma)\rangle
=-\langle \sigma_0\rangle + \langle f(s_0)\rangle
\end{equation}
or
\begin{equation}
\frac{d}{dt}\langle \sigma_0\rangle =
-\langle \sigma_0\rangle + A\langle\sigma_i\rangle
+B\langle\sigma_i\sigma_j\sigma_k\rangle
+C\langle\sigma_i\sigma_j\sigma_k\sigma_\ell\sigma_n\rangle
\end{equation}
where translation invariance has been used.
Next, we use the simplest mean-field approximation, for
which $\langle\sigma_i\sigma_j\sigma_k\rangle=m^3$
and $\langle\sigma_i\sigma_j\sigma_k\sigma_\ell\sigma_n\rangle=m^5$,
where $\langle\sigma_i\rangle=m$, to write
\begin{equation}
\frac{dm}{dt} = -(1 - A)m + B m^3 + C m^5.
\end{equation}

%-------------
\begin{figure}
\centering
\epsfig{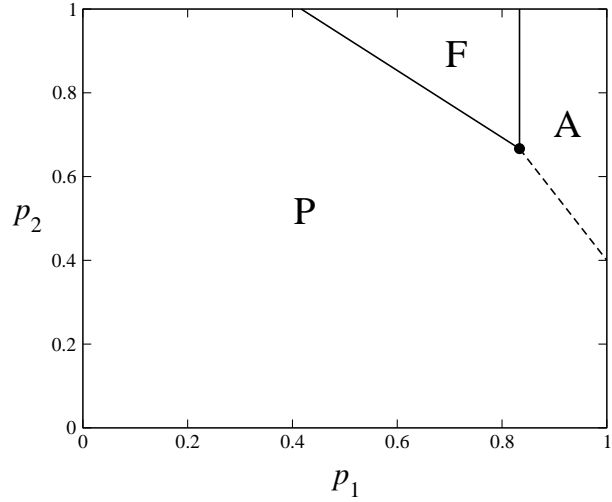}
\caption{Phase diagram in the plane $(p_1,p_2)$ according to
mean-field approximation. The full circle, located at $(p_1,p_2)=(5/6,2/3)$,
corresponds to the voter model. The phases are:
paramagnetic (P), ferromagnetic (F) and absorbing (A). 
The solid lines are continuous phase transitions and the
dashed line is a discontinuous phase transition.}
\label{diagmf}
\end{figure}
%-----------

Here, we are interested in the case $p_0=1$ which 
is equivalent to $1-A=B+C$ so that
\begin{equation}
\frac{dm}{dt} = -(B+C)m + B m^3 + C m^5,
%= m(- B - C - Cm^2)(1 - m^2).
% ver eq. (2) de hammal05
\label{20}
\end{equation}
which is identical to the deterministic part of the
Langevin equation considered in reference \cite{hammal05}.
In the stationary state, the possible solutions are as follows.
a) $m=0$, corresponding to the paramagnetic (P) phase;
it is stable as long as $B+C>0$, which is equivalent to $12p_1+15p_2< 20$.
b) $m=\pm1$, corresponding the absorbing (A) phase;
it is stable as long as $B+2C<0$, which is equivalent to $p_1>5/6$.
c) $m\neq0$ and $m\neq\pm1$, corresponding to the ferromagnetic (F) phase.
The phase diagram in the plane $(p_1,p_2)$ is shown in figure \ref{diagmf}.
The P-F line is given  by $B+C=0$, or  $12p_1+15p_2=20$,
and the F-A line is given by $B+2C=0$, or $p_1=5/6$.
The point corresponding to the voter model is located at the point where
the two lines meet, that is, at $B=C=0$, or $p_1=5/6$ and $p_2=2/3$.
The P-F and F-A transitions are continuous whereas the P-A transition
is discontinuous. 

The discontinuous P-A transition line shown in figure \ref{diagmf}
was obtained as follows. We start by writing equation (\ref{20})
in the form $dm/dt=-\partial f/\partial m$ where
$f=(B+C)\,m^2/2-Bm^4/4-Cm^6/6+K$. Interpreting $f(m)$ as a
free energy then the discontinuous transition occurs when
$f(0)=f(\pm1)$, which gives $3B+4C=0$ or $5p_2+8p_1=10$. 

Inside the F phase, the magnetization $m$ is
given by
\begin{equation}
m = \left(\frac{15p_2+12p_1-20}{15p_2-12p_1}\right)^{1/2}.
\end{equation}
Near the critical line PF, $p_1=p_{\rm FP}=(20-15p_2)/12$, we may write
\begin{equation}
m = A(p_1-p_{\rm FP})^{1/2},
\qquad\qquad 
A = \left[\frac52\left(p_2-\frac23\right)\right]^{-1/2}
\end{equation}
so that the amplitude $A$ diverges with an exponent $1/2$
as one approaches the voter point.
Analogously, near the FA transition, $p_1=p_{\rm FA}=5/6$,
the order parameter $\rho$ behaves as
\begin{equation}
\rho = B(p_{\rm FA}-p_1),
\qquad\qquad
B = \left[\frac54\left(p_2-\frac23\right)\right]^{-1}
\end{equation}
so that the amplitude $B$ diverges with an exponent $1$
as one approaches the voter point.
It is worth to notice that the size $\ell=p_{\rm FA}-p_{\rm PA}$
of the ferromagnetic phase, given by
\begin{equation}
\ell = \frac54\left(p_2-\frac23\right)
\end{equation}
vanishes linearly as one approaches the voter point.

%-------------
\begin{figure}
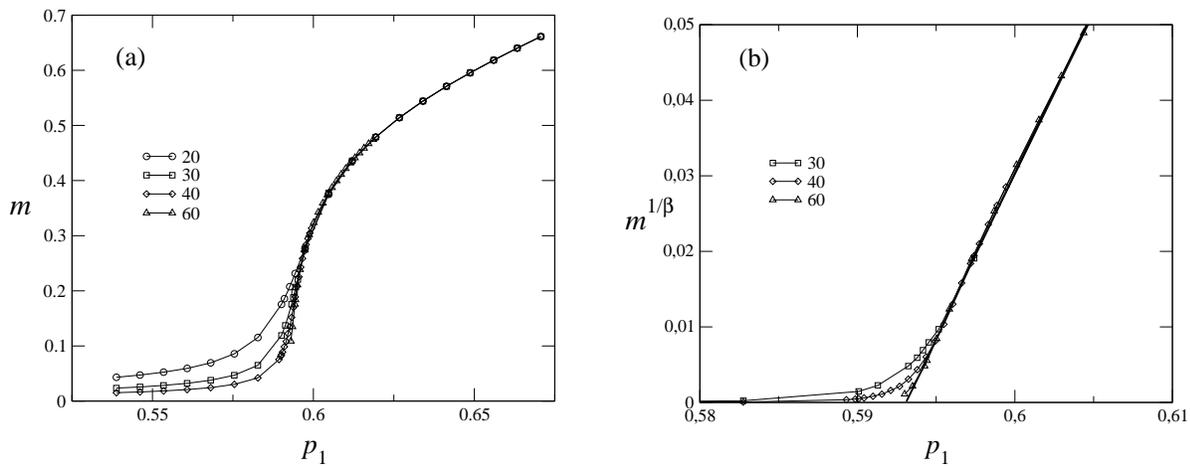

\centering
\epsfig{file=mag13D.eps,width=7.3cm}
\hfill
\epsfig{file=mag13Dexp.eps,width=7.5cm}
\caption{(a) Magnetization $m$  versus
$p_1$ along $p_2=1$. The transition occurs at $p_{I}=0.5930$
as can be inferred from the plot (b) of $m^{1/\beta}$ versus $p_1$,
where $\beta=0.326$.}
\label{mag13D}
\end{figure}
%-----------

%----------------------------------------------------------
\section{Simulations}

The results obtained from numerical simulations on the cubic and
triangular lattice allow us to conclude that the topology of the
phase diagram is the same as the one obtained from the mean-field
approximation. 
The numerical simulations on these lattices were
carried out as follows. At each time step a site of the lattice
is chosen at random and updated according to the transition  probabilities
in table \ref{trans}, restricted to the case $p_0=1$.
The averages of several quantities were 
obtained in the stationary state. Defining $s=(1/N)\sum_i\sigma_i$
where $N$ is the number of sites, 
then the quantities of interest to analyze the ferromagnetic-paramagnetic
line are the magnetization
\begin{equation}
m = \langle|s|\rangle,
\end{equation}
the susceptibility,
\begin{equation}
\chi = N\{ \langle s^2\rangle - \langle|s|\rangle^2 \},
\end{equation}
the Binder cumulant 
\begin{equation}
U = 1-\frac{\langle s^4\rangle}{3\langle s^2\rangle^2}.
\end{equation}

\begin{figure}
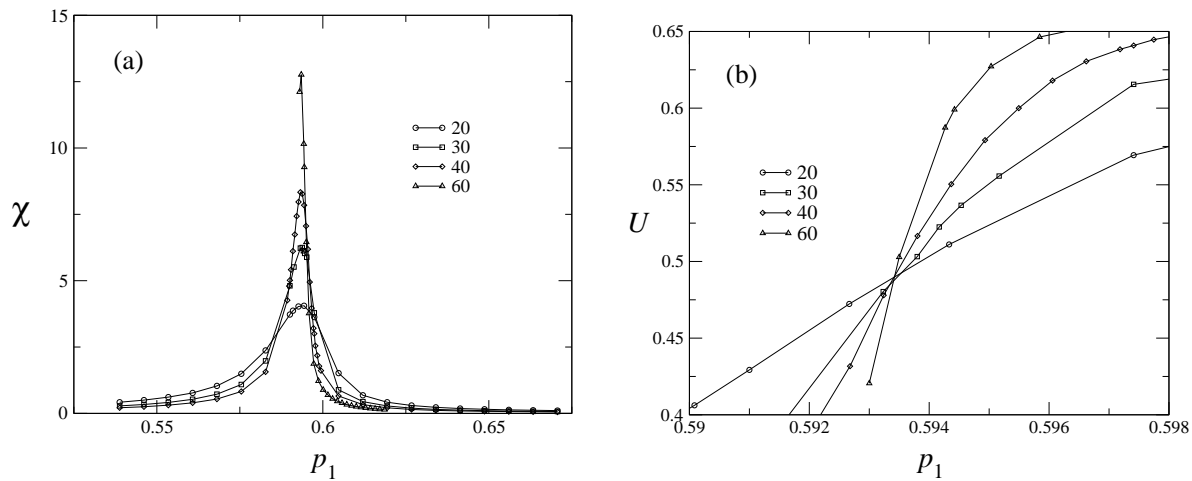

\centering
\epsfig{file=susc13D.eps,width=7.5cm}
\hfill
\epsfig{file=bin13D.eps,width=7.5cm}
\caption{(a) Susceptibility $\chi$ and (b) Binder cumulant $U$
versus $p_1$ along $p_2=1$. The transition occurs at $p_{I}=0.5933$.}
\label{susc13D}
\end{figure}

Examples of these quantities, for the cubic lattice with
$N=L^3$ sites, calculated along the line $p_2=1$,
are shown in figures \ref{mag13D} and \ref{susc13D}.
At the P-F transition point, the Binder cumulant $U$ is a
constant, a result that allows to located the P-F transition line. 
The critical behavior of $m$ and $\chi$ allows
to say that, with respect to the ferromagnetic-paramagnetic
transition, the model belongs to the universality class of the Ising model.

%-------------
%-----------

\begin{figure}
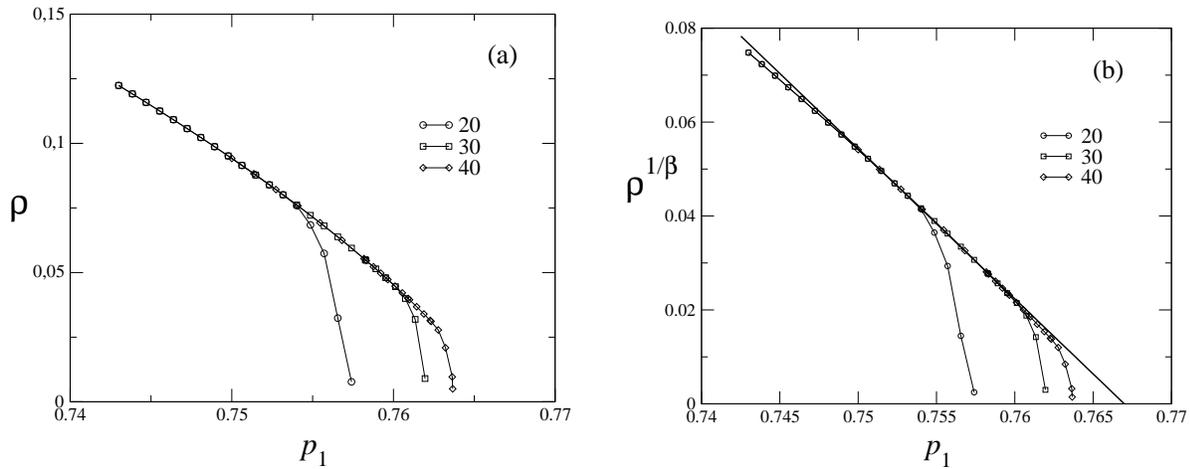

\centering
\epsfig{file=rho13D.eps,width=7.5cm}
\hfill
\epsfig{file=rho13Dexp.eps,width=7.5cm}
\caption{(a) Order parameter $\rho$ of the F-A transition
as a function of $p_1$ for $p_2=1$, 
The critical point occurs at $p_{DP}=0.7670$
as can inferred from the plot (b) where $\beta=0.81$}
\label{rho13D}
\end{figure}

To analyze the ferromagnetic-absorbing transition, where
the ferromagnetic phase is identified as the active phase,
we define the quantity $\rho = 1-m$, which is the
order parameter that characterizes the transition
from active to absorbing.  
In figure \ref{rho13D}, we show $\rho$ versus $p_1$ along the
line $p_2=1$. The location of the transition
is obtained by the extrapolation $\rho\to0$, what allows us to draw
the F-A line. The critical behavior shows that
with respect to this transition, the
model belongs to direct percolation universality class.

\begin{figure}
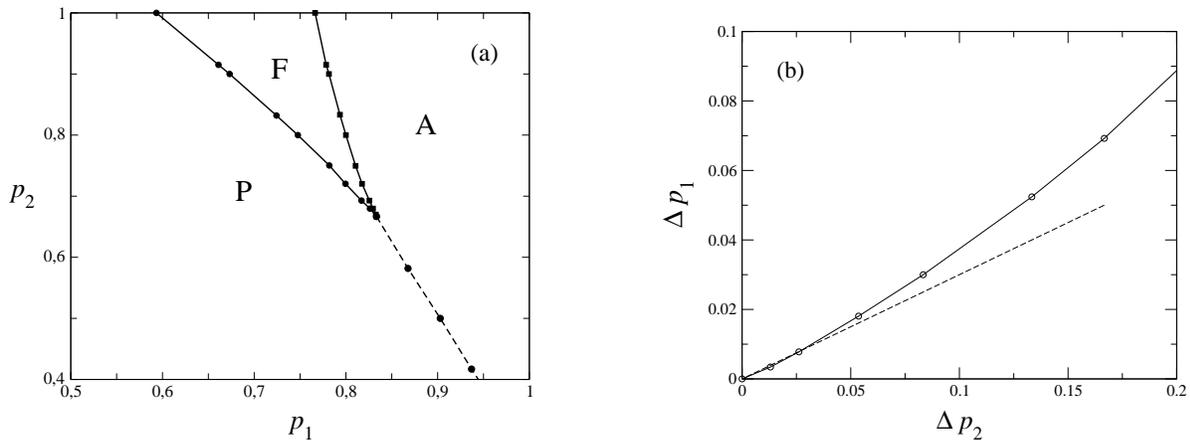

\centering
\epsfig{file=diagcubx.eps,width=7cm}
\hfill
\epsfig{file=larg.eps,width=7cm}
\caption{(a) Phase diagram in the plane $(p_1,p_2)$ according to
simulations on a cubic lattice.
The phases are:
paramagnetic (P), ferromagnetic (F) and absorbing (A). 
The solid lines are continuous phase transitions and the
dashed line is a discontinuous phase transition.
The three lines meet at the point $(p_1,p_2)=(5/6,2/3)$,
corresponding to the voter model. (b) Linear size 
$\Delta p_1=p_{\rm FA}-p_{\rm PF}$
of the ferromagnetic phase as a function of the distance
from the voter point $\Delta p_2=p_2-2/3$. The dashed straight
line shows that the data points have a finite slope at the origin.
}

\label{diagcub}
\end{figure}

Using the data of the P-F and F-A transitions obtained for several
values of $p_2$ we draw the phase diagram of figure \ref{diagcub}a.
As expected the two lines, P-F and F-A, meet at the voter point
located at $p_1=5/6$ and $p_2=2/3$. In figure \ref{diagcub}b
we show the size $\Delta p_1 = p_{DP}-p_{I}$ of the F phase
as a function of the distance from the voter point $\Delta p_2=p_2-2/3$.
As can be seen in this figure, the data points has a finite
slope at the origin which amounts do say that the scaling relation
\begin{equation}
\Delta p_1 \sim (\Delta p_2)^\phi,
\end{equation}
is fulfiled with $\phi=1.0$.

%gives $\phi=1.13(2)$. 
%This result is in fairly agreement
%with the result $\phi=1.09(2)$ obtained from the ratio
%$\phi=\nu_{\rm I}/\nu_{\rm DP}$, where $\nu_{\rm I}=0.630$
%%Pelisseto e Vicari, nosso livro p. 262
%and $\nu_{\rm DP}=0.58$
%%nosso livro p. 282
%in three dimensions.

We have also determined the nature of the phase transition
between the P and A phases. As we cross the transition line
the order parameter jumps from zero, which is the value of $m$ within
the paramagnetic phase, to the maximum
value $m=1$, which is the value of $m$ inside the absorbing phase.
The susceptibility $\chi$ increases as one approaches the
transition line from the paramagnetic phase and reaches a
finite value $\chi_0(\Delta p_2)$ which depends on the
distance $\Delta p_2=p_2-2/3$ from the voter point. As
one approaches the voter point $\chi_0$ increases without
limits and diverges with an exponent $\gamma=1$.
Notice that, according to our numerical results for the susceptibility
obtained along $p_2=2/3$, the susceptibility $\chi$ diverges as one
approaches the voter point from the paramagnetic phase
with the same exponent $\gamma=1$.

%-------------
%-----------

\begin{figure}
\centering
\epsfig{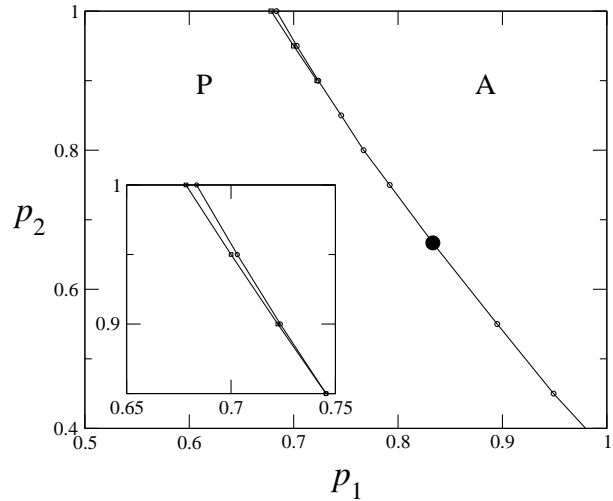}
\caption{Phase diagram in the plane $(p_1,p_2)$ according to
simulations on a triangular lattice.
The full circle, located at $(p_1,p_2)=(5/6,2/3)$,
corresponds to the voter model. The phases are:
paramagnetic (P), ferromagnetic and absorbing (A).
The inset shows an enlargement around showing the ferromagnetic
phase between the P-F and F-A transition lines.}
\label{diagtri}
\end{figure}

We have also studied the model defined on a triangular lattice
by using the same methods. As can be seen in figure \ref{diagtri},
the phase diagram also displays the three phases found in the 
cubic lattice. However, the ferromagnetic phase is very narrow,
becoming very difficult to extract the crossover exponent
but the two lines, PF and FA, seems to meet tangentially
in consistency with an exponent $\phi=1$ with logarithm corrections
as we will argue below.

%-------------
%-----------

%-------------
%-----------

\section{Discussion}

We have studied the phase diagram of a system belonging to
the Ising and DP universality classes. The phase diagram
in the two parameters has an Ising critical line
separating the paramagnetic and the ferromagnetic phase,
a DP critical line separating the ferromagnetic phase
and the absorbing phase,
and a discontinuous phase transition line separating
the paramagnetic and the absorbing phase. The three lines
meet at the voter point in such as way that the size
of the ferromagnetic phase vanishes as one approaches
the voter point with a crossover exponent $\phi$.
%The exponent $\phi$ is found to be equal do $\phi=1$ within
%the mean-field theory and $\phi=1.0$ in three dimensions.
The crossover exponent should be understood as the ratio
between the two exponents related to the two relevant scaling fields
$\varepsilon$ and $r$ around the voter point.
Above the critical dimension $d_c$ of the voter model,
these two exponents do not depend on dimension so that the
same should happen to the crossover exponent.
Taking into account that the upper critical dimension of the voter
critical point is $d_c=2$, we expect the crossover exponent $\phi$
to have the same value for $d\geq2$ with possible logarithm corrections
when $d=2$. The results coming from mean-field theory gives
$\phi=1$ which allows to conclude that $\phi=1$ for $d\geq2$, a result 
confirmed by our calculations in $d=3$. As to the two-dimensional
case, the size of the ferromagnetic phase is too narrow
for a precise numerical calculation but is consistent
with the result $\phi=1$ with logarithm corrections.

The results for the susceptibility around the Ising
and DP lines are in accordance with these two
universality classes. Concerning the discontinuous
transition from the paramagnetic to the absorbing
transition, we found that the susceptibility is
finite at the transition line but diverges as one
approaches the voter point. 

It would be very interesting if the picture that we
have drawn concerning the critical behavior
of systems belonging to the Ising and DP universality classes
could be confirmed by a renormalization group calculation
by means of a Langevin equation formulation.

%%--------------------------------------------------------
%\section*{Acknowledgment}

%We wish to acknowledge C. for his helpful comments.

%----------------------------------------------------------

\end{document}